\pdfoutput=1
\documentclass[a4paper,reprint,twocolumn,notitlepage,aip,footinbib]{revtex4-2}

\usepackage[left=1.5cm,right=1.5cm,top=1cm,bottom=1.5cm,includeheadfoot]{geometry}

\usepackage{tgtermes} 
\usepackage{tgheros} 
\usepackage[T1]{fontenc}

\usepackage{amssymb}
\usepackage{amsmath}
\usepackage{amsthm}
\usepackage[dvipsnames]{xcolor}
\usepackage{soul}
\usepackage{float}
\usepackage{multirow}
\usepackage{enumerate}
\usepackage{stmaryrd}

\usepackage{hyperref}
\usepackage[dvipsnames]{xcolor}
\usepackage{bm} 
\usepackage[normalem]{ulem}
\usepackage{tikz}

\newcommand{\rmi}{\mathrm{i}}
\newcommand{\R}{\mathbb{R}}
\newcommand{\rr}{\mathbf{r}}
\newcommand{\dd}{\mathrm{d}} 
\newcommand{\eps}{\varepsilon}
\newcommand{\rightder}{\partial^+}
\newcommand{\leftder}{\partial^-}
\DeclareMathOperator{\trace}{Tr}

\newcommand{\EHxc}{E_\mathrm{Hxc}}
\newcommand{\Exc}{E_\mathrm{xc}}
\newcommand{\Ex}{E_{\rm x}}
\newcommand{\Ec}{E_{\rm c}}
\newcommand{\EHartree}{E_{\rm H}}

\newcommand{\vxc}{v_\mathrm{xc}}
\newcommand{\vx}{v_\mathrm{x}}
\newcommand{\vH}{v_\mathrm{H}}

\newcommand{\vHxc}{v_\mathrm{Hxc}}
\newcommand{\vc}{v_\mathrm{c}}

\begin{document}

\title{Exchange-only virial relation from the adiabatic connection}
\author{Andre Laestadius}
\email[Electronic address:\;]{andre.laestadius@oslomet.no}
\affiliation{Department of Computer Science, Oslo Metropolitan University, Norway}
\affiliation{Hylleraas Centre for Quantum Molecular Sciences, Department of Chemistry, University of Oslo, Norway}

\author{Mih\'aly A. Csirik}
\affiliation{Department of Computer Science, Oslo Metropolitan University, Norway}
\affiliation{Hylleraas Centre for Quantum Molecular Sciences, Department of Chemistry, University of Oslo, Norway}

\author{Markus Penz}
\affiliation{Department of Computer Science, Oslo Metropolitan University, Norway}
\affiliation{Max Planck Institute for the Structure and Dynamics of Matter and Center for Free-Electron Laser Science \& Department of Physics, Hamburg, Germany}

\author{Nicolas Tancogne-Dejean}
\affiliation{Max Planck Institute for the Structure and Dynamics of Matter and Center for Free-Electron Laser Science \& Department of Physics, Hamburg, Germany}

\author{Michael Ruggenthaler}
\affiliation{Max Planck Institute for the Structure and Dynamics of Matter and Center for Free-Electron Laser Science \& Department of Physics, Hamburg, Germany}
\affiliation{The Hamburg Center for Ultrafast Imaging, Hamburg, Germany}

\author{Angel Rubio}
\affiliation{Max Planck Institute for the Structure and Dynamics of Matter and Center for Free-Electron Laser Science \& Department of Physics, Hamburg, Germany}
\affiliation{Center for Computational Quantum Physics, Flatiron Institute, New York, USA}
\affiliation{The Hamburg Center for Ultrafast Imaging, Hamburg, Germany}

\author{Trygve Helgaker}
\affiliation{Hylleraas Centre for Quantum Molecular Sciences, Department of Chemistry, University of Oslo, Norway}

\begin{abstract}
The exchange-only virial relation due to Levy and Perdew is revisited. 
Invoking the adiabatic connection, we introduce the exchange energy in terms of the right-derivative of the universal density functional w.r.t.\ the coupling strength $\lambda$ at $\lambda=0$. This agrees with the Levy--Perdew definition of the exchange energy as a high-density limit of the full exchange-correlation energy.  
By relying on $v$-representability for a fixed density at varying coupling strength, we prove an exchange-only virial relation without an explicit local-exchange potential. Instead, the relation is in terms of a limit ($\lambda \searrow 0$) involving the exchange-correlation potential $\vxc^\lambda$, which exists by assumption of $v$-representability. On the other hand, a local-exchange potential $\vx$ is not warranted to exist as such a limit.
\end{abstract}

\maketitle

\section{Introduction}
\label{sec:intro}

Despite the long history of density-functional theory (DFT) and research concerning its foundation, 
DFT still has many mathematical challenges remaining.\cite{teale2022round-table,Penz-et-al-HKreview-partI,Wrighton-et-al-2023,Crisostomo2023}
Addressing these contribute to the development of improved approximate functionals and enhance the overall understanding of DFT. The aim of this work is to investigate the exchange-only virial relation of Perdew and Levy from a  mathematical standpoint, motivated by some recent developments in mathematical DFT.\cite{Lammert2007,Kvaal2014,KSpaper2018,Penz-et-al-ES-2023,other-special-edition-paper} It is with great enthusiasm we submit this contribution to the special issue honoring the great achievements of John Perdew in the field of quantum chemistry in general and DFT in particular. 

Exact constraints play an important role in the development and testing of density-functional approximations,\cite{LevyPerdewPhysRevA1985,RevModPhys.61.1989,HuiLevy1990,LevyPerdewPhysRevB1993,LevyPerdewIJQC1994} bearing in mind that semiempirical functionals can fail outside their training set.\cite{Blaha1999}
These constraints are based on conditions that the exact exchange--correlation functional $E_\mathrm{xc}[\rho]$ or its constituent exchange $\Ex[\rho]$ and correlation
$\Ec[\rho]$ parts satisfy. 
For example, the second-order gradient expansion, aimed at improving on the local density approximation (LDA), can perform worse than LDA for real systems because LDA satisfies certain exact constraints that finite-order gradient expansions break.\cite{GunnarssonLundqvist1976,PBEprl1996,PBW1996}
In the generalized gradient approximation (GGA), these constraints are restored, leading to an overall better performance of the functionals.

One such constraint for GGA~\cite{Engel1993,Borlido2020} follows from a physically appealing formula that arises as a consequence of the general virial theorem of quantum mechanics. In particular,
\citet{LevyPerdewPhysRevA1985} have proposed the exchange-only virial relation
\begin{align}\label{eq:Ex-VR}
    \Ex[\rho] = - \int \rho(\rr) \, \rr\cdot \nabla \vx(\rr) \, \dd \rr.
\end{align}
Relying on a different work of \citet{LevyPerdewIJQC1994}, we \emph{define} the exchange energy as the high-density limit of the exchange--correlation energy in the manner 
\begin{equation}\label{eq:def-of-Ex}
    \Ex[\rho] = \lim_{\gamma \to \infty} \frac 1 \gamma \Exc[\rho_\gamma] ,
\end{equation}
where the scaled density is given by $\rho_\gamma(\rr) = \gamma^3 \rho(\gamma\rr)$.
The exchange--correlation energy $\Exc[\rho]$ is defined in terms of the universal density functional $F[\rho]$ in the usual manner; see Eq.~\eqref{eq:Exc-def}. Noting that $(\rho_\gamma)_\mu = \rho_{\gamma\mu}$, it follows directly from Eq.\,\eqref{eq:def-of-Ex} that the exchange energy obeys the well-known\cite{LevyPerdewPhysRevA1985,kaplan2023exact-constraints} exact constraint $\Ex[\rho_\gamma] = \gamma \Ex[\rho]$.

There exists an alternative, equivalent definition for $\Ex[\rho]$ that employs the adiabatic connection in DFT. This extremely useful concept relates not only the non-interacting system (with interaction strength $\lambda=0$) with the fully interacting one ($\lambda=1$) for a fixed, given density $\rho$, as it is customary in Kohn--Sham DFT, but also reveals the properties of the functionals for all intermediate values of the coupling strength $\lambda$. In the following, 
we study $F[\rho]$ as a function the interaction strength for a fixed, given density $\rho$,
using the notation $\lambda \mapsto F(\lambda)$; see Eq.~\eqref{eq:F-def}. The theory of convex analysis then provides multiple tools to study the adiabatic connection. Already from the definition of the universal density functional, many practical properties can be derived when considering it as a function of $\lambda$. The intuitive understanding of the exchange energy in this setting is then as the linear component of $\lambda \Exc(\lambda) = F(\lambda) - F(0) - \lambda\EHartree$, where $\EHartree$ is the Hartree energy (see Eq.~\eqref{eq:Hartree}).

Here, all derivations will be performed for general, mixed $N$-electron states, thus allowing for degeneracy at all coupling strengths, with the overall assumption of a $v$-representable density $\rho$ at all coupling strengths $\lambda \in \mathbb R$ (a property henceforth called $v^\lambda$-representability). This comes as a form of minimal assumption, since, without $v^\lambda$-representability, the selected density cannot be assigned to a valid ground state of the system under consideration. 
On the other hand, it cannot be assumed that the universal density functional is functionally differentiable with respect to the density, nor does this need to hold for its individual parts such as the exchange functional. Equation~\eqref{eq:Ex-VR} would then be the direct consequence of a functionally differentiable $\Ex[\rho]$ (see App.~B of \citet{other-special-edition-paper} in the same special issue of this journal), but, 
unfortunately, the universal density functional is everywhere discontinuous.\cite{Lammert2007}

The principal aim of this paper to find a version of the exchange virial relation that holds without assuming functional differentiability. 
We shall find that this is indeed possible using a limit procedure 
and the main result, derived in Section~\ref{sec:xVR}, is 
\begin{equation}\label{eq:Ex-VR-main-result}
    \Ex = -\lim_{\lambda\searrow 0}\int \rho(\rr) \,\rr \cdot \nabla \vxc^\lambda(\rr) \,\dd \rr,
\end{equation}
as opposed to Eq.~\eqref{eq:Ex-VR}. Here, $\vxc^\lambda(\rr)$ is the corresponding exchange-correlation potential for interaction strength $\lambda$. 
However, this procedure does not provide us with a well-defined local-exchange potential $v_{\rm x}(\rr) = \lim_{\lambda\searrow 0} \vxc^\lambda(\rr)$, as one would perhaps expect. Indeed, a general local-exchange potential need not exist, a finding that agrees with the force-based approach to DFT, where an additional vector potential is introduced to fulfill the exchange-only virial relation.\cite{other-special-edition-paper} This also opens the possibility to a \emph{non-local} limit potential.

We begin by giving the basic setting and some useful relations for the expectation values of the main components of the Hamiltonian. In atomic units, we  consider a nonrelativistic system of $N$ spinless particles in $\R^3$, for which the Hamiltonian at coupling strength $\lambda \in \R$ is given by
\begin{equation}\label{eq:Hamiltonian}
    \hat H^\lambda[v] = \hat T + \lambda \hat W + \sum_{j=1}^N v(\rr_j) ,
\end{equation}
where $\hat T$ and $\hat W$ denote the kinetic and two-particle interaction operators, respectively:
\begin{equation}\label{eq:T-W-def}
    \hat  T = - \frac{1}{2} \sum_{j=1}^N \nabla_{\rr_j}^2, \quad \hat W = \sum_{1\leq j<k\leq N} \frac{1}{|\rr_j-\rr_k|} .
\end{equation}
Both operators are positive definite. 
We can bound the expectation value of $\hat W$ in terms of $\hat T$, for instance, by using the Hardy-, and the Cauchy--Schwarz inequalities, we have
\begin{equation}\label{eq:W-T-bound}
    \trace \hat W \Gamma \leq C_N \sqrt{\trace \hat T  \Gamma}
    \leq \frac{C_N}{2} (1 + \trace\hat T \Gamma )
\end{equation}
for some $N$-dependent constant $C_N>0$.

If $v(\rr)$ is such that it admits a ground state with density $\rho(\rr)$ then the full virial relation for such a system can be derived  from an application of the Ehrenfest theorem for mixed states. For a ground state $\Gamma$ it states that $\rmi \trace [\hat H, \hat A] \Gamma = \partial_t \trace \hat A \Gamma = 0$ and we can apply this to the operator $\hat A = \sum_{j=1}^N \rr_j\cdot\nabla_j$. This gives~\cite{hirschfelder1960classical,theophilou2020virial}
\begin{equation}\label{eq:full-VR}
    2\trace \hat T \Gamma + \lambda \trace \hat W \Gamma = \int \rho(\rr) \,\rr \cdot \nabla v(\rr) \, \dd \rr,
\end{equation}
where $\rho \mapsfrom \Gamma$ is the ground-state density associated 
with the mixed state $\Gamma$.
This expression forms the basis for our discussion of the virial relation in DFT.

\section{Coupling-strength density functionals}
\label{sec:coupling-strength-functionals}

Central to DFT is the universal density functional, which, for a given coupling strength $\lambda \in \R$ and density $\rho$ is defined as the constrained search over all density matrices that yield a given one-particle density $\rho$,
\begin{equation}\label{eq:F-def}
    F(\lambda) = \inf_{\Gamma\mapsto \rho} \trace (\hat T +\lambda  \hat W) \Gamma.
\end{equation}
Since the density $\rho$ is kept fixed in our discussion, we omit it 
as an argument to the functional. The only exception is when the scaled density $\rho_\gamma$ rather than $\rho$ is considered, always at $\lambda=1$, in which case we write $F[\rho_\gamma]$. 
By Corollary~4.5 in \citet{Lieb1983}, for all `physical', $N$-representable densities $\rho$ (non-negative, normalized, and of finite kinetic energy), there exists a minimizing (possibly non-unique) density matrix $\Gamma(\lambda)$ in Eq.~\eqref{eq:F-def} such that $\Gamma(\lambda) \mapsto \rho$ and
\begin{equation}\label{eq:F-Gamma}
    F(\lambda) = \trace (\hat T +\lambda  \hat W) \Gamma(\lambda).
\end{equation}
Although this corollary is formulated with $\lambda\geq 0$, it can easily be extended to all $\lambda \in \mathbb R$.~\footnote{In the proof of Theorem 4.4 in~\citet{Lieb1983}, we replace $h^2 = \hat T + \lambda \hat W +1$ by $h^2 = \hat T + \lambda \hat W +1 + M_\lambda$, where $-M_\lambda$ is the lower bound of the combined operator $\hat T + \lambda \hat W$ given by Kato--Rellich theorem~\cite{reed-simon-2}.}
Existence of a minimizer in Eq.~\eqref{eq:F-def} ensures that $F(\lambda)$ is finite for all $\lambda \in \R$. 

Since the minimizing density matrix $\Gamma(\lambda) \mapsto \rho$ in Eq.\,\eqref{eq:F-Gamma} is in general not unique, it is \emph{a priori} unclear how to decompose the function $\lambda\mapsto F(\lambda)$ uniquely into a kinetic and an interaction contribution. For every choice of minimizing $\Gamma(\lambda)\mapsto\rho$, however, we \emph{do} have the decomposition
\begin{equation}
F(\lambda) = T(\Gamma(\lambda)) + \lambda W(\Gamma(\lambda))
\end{equation}
with
\begin{alignat}{3}
    &T(\Gamma(\lambda)) = \trace{\hat T \Gamma(\lambda)}, \label{eq:T-lambda-formula} \\
    &W(\Gamma(\lambda)) = \trace{\hat W \Gamma(\lambda)}. \label{eq:W-lambda-formula}
\end{alignat}
As we shall now show, this decomposition becomes unique under the assumption of $v$-representability of the density $\rho$ at interaction strength $\lambda$. Hence, instead of $T(\Gamma(\lambda))$ and $W(\Gamma(\lambda))$ we may write $T(\lambda)$ and $W(\lambda)$ and so
\begin{equation}
F(\lambda) = T(\lambda) + \lambda W(\lambda)
\label{eq:FTW}
\end{equation}
holds. 

To demonstrate uniqueness, we first note that $v$-representability for $\rho$ means that we can find a potential $v^\lambda(\rr)$ such that there is a ground state $\Gamma(\lambda) \mapsto \rho$ for the Hamiltonian $H^\lambda[v^\lambda]$. 
By the virial theorem, Eq.~\eqref{eq:full-VR}, we then have
\begin{equation}
    2 T(\Gamma(\lambda)) + \lambda W(\Gamma(\lambda)) = \int \rho(\rr) \,\rr \cdot \nabla v^\lambda(\rr) \,\dd \rr.
\end{equation}
Subtracting $F(\rho)$ as given in Eq.\,\eqref{eq:F-Gamma} from both sides of this equation and using Eq.\,\eqref{eq:FTW}, we obtain
\begin{equation}\label{eq:proof-structure}
    T(\Gamma(\lambda)) = \int \rho(\rr) \,\rr \cdot \nabla v^\lambda(\rr) \,\dd \rr - F(\lambda),
\end{equation}
whose right-hand side is uniquely determined by the density, i.e., it is independent of the choice of $\Gamma(\lambda)$. 
For $F(\lambda)$ this is clear while for $\nabla v^\lambda(\rr)$ it follows from the Hohenberg--Kohn theorem~\cite{Hohenberg1964}, which states that $v^\lambda(\rr)$ is determined uniquely up to a scalar by the ground-state density. Hence $T(\lambda):=T(\Gamma(\lambda))$ is well-defined (for all $\lambda)$ and $W(\lambda):=W(\Gamma(\lambda))$ is also well-defined (for $\lambda\neq 0$) via $\lambda W(\lambda) = F(\lambda) - T(\lambda)$ implying that
\emph{the universal density functional has a unique decomposition into kinetic-, and interaction contributions at each interaction strength where the density is $v$-representable.} 
Another possibility to achieve well-defined functions $T(\lambda)$ and $W(\lambda)$ is discussed in  Section~\ref{sec:remarks}. 
We will consider the case $W(0)$ later.

In the pure-state formulation, the functionals are  defined as expectation values with $\Psi(\lambda) \mapsto \rho$ as the minimizer in the constrained-search functional. It is  possible that this minimizing state is then uniquely given, and so the functionals are already defined unambiguously, but this does not seem to be guaranteed. Consequently,  by the same argument as above, we rely on \emph{pure-state} $v^\lambda$-representability to make sure that $\langle \Psi(\lambda) | \hat T | \Psi(\lambda) \rangle$ and $\langle \Psi(\lambda) | \hat W | \Psi(\lambda) \rangle$ have values that only depend on $\rho$ for $\lambda>0$. Only if $\Psi(0)$ is determined uniquely -- for example, if $\rho$ is noninteracting pure-state $v$-representable by a nondegenerate ground state -- can also $W(0) = \langle \Psi(0) | \hat W | \Psi(0) \rangle$ be given, as it is the case in most presentations on the subject. Here, we aim to avoid such restrictions, only demanding \emph{ensemble} $v$-representability of $\rho$, defining $W(0)$ by a limit procedure that is laid out in Section~\ref{sec:diffprop}.

\section{Basic properties of the adiabatic connection}
\label{sec:prop}

Recall that by our assumption of $v^\lambda$-representability of the considered density $\rho$ and following the discussion in Sec.~\ref{sec:coupling-strength-functionals}, the individual parts $T(\lambda)$ and $W(\lambda)$ of the universal density functional $F(\lambda)$ are uniquely defined. 
We  proceed by collecting important properties of the function $F(\lambda)$ that stem directly from its definition by Eq.~\eqref{eq:F-def}. Firstly, we relate $F(\lambda)$ with $\lambda=1/\gamma > 0$ to the scaled density as~\cite{LevyPerdewIJQC1994}
\begin{equation}\label{eq:F-scaling}
\begin{aligned}
\gamma^2 F(1/\gamma) = &\inf_{\Gamma \mapsto \rho} \trace{(\gamma^2 \hat T + \gamma \hat W) \Gamma} \\
=& \inf_{\Gamma \mapsto \rho} \trace{(\hat T + \hat W) \Gamma_\gamma} \\
=& \inf_{\Gamma_\gamma \mapsto \rho_\gamma} \trace{(\hat T + \hat W) \Gamma_\gamma} 
= F[\rho_\gamma],
\end{aligned}
\end{equation}
where $\Gamma_\gamma = \gamma^{3N}\Gamma(\gamma \cdot)$ is the coordinate-scaled density matrix that yields precisely $\rho_\gamma$. 

Importantly, $F(\lambda)$ is concave, as follows easily from the superadditivity of the infimum -- namely, for each $\lambda, \lambda' \in \R$ and $s\in [0,1]$, we have
\begin{equation}
\begin{aligned}
    &F(s\lambda + (1-s)\lambda') = \inf_{\Gamma\mapsto \rho} \trace (\hat T +(s\lambda + (1-s)\lambda') \hat W) \Gamma \\
    &\geq s\inf_{\Gamma\mapsto \rho} \trace (\hat T + \lambda \hat W) \Gamma + (1-s)\inf_{\Gamma\mapsto \rho} \trace (\hat T + \lambda' \hat W) \Gamma\\
    &= s F(\lambda) + (1-s) F(\lambda').
\end{aligned}
\end{equation}

Moreover, since the operator $\hat W$ is positive definite, we have that $W(\lambda)$ is strictly positive and thus $F(\lambda)$ is strictly monotonically increasing.
Further, it holds that 
\begin{equation}
\begin{aligned}
    F(\lambda') &= T(\lambda') + \lambda' W(\lambda') \\
    &\leq T(\lambda) + \lambda' W(\lambda) = F(\lambda) + (\lambda'-\lambda)W(\lambda),
\end{aligned}
\end{equation}
where the inequality stems from the fact that on the right-hand side $T(\lambda)$ and $W(\lambda)$ are constructed from $\Gamma(\lambda)$, which in general is not a minimizer for $F(\lambda')$. From
\begin{equation}
    F(\lambda') \leq F(\lambda) + (\lambda'-\lambda)W(\lambda),
\end{equation}
we see that $W(\lambda)$, $\lambda \neq 0$, is an element of the superdifferential of $F(\lambda)$. The individual elements of the set-valued superdifferential are called supergradients and they form tangents lying entirely on or above the respective concave function.
Any choice of supergradients for different $\lambda$ must be monotonically decreasing since $F(\lambda)$ is concave~\cite{niculescu-persson-book}, so this also holds for $W(\lambda)$. Then, by a generalization of the fundamental theorem of integral calculus~\cite{niculescu-persson-book}, we can write $F(\lambda) = F(0) + \int_0^\lambda f(\mu)\, \dd \mu$ for any choice of supergradient $f(\mu)$ of $F(\mu)$. By choosing $f(\mu) = W(\mu)$ we get, in particular, 
\begin{equation}\label{eq:F-fundamental-th}
    F(\lambda) = F(0) + \int_{0}^\lambda W(\mu) \, \dd \mu.
\end{equation}
This result also follows as a direct consequence of the Hellmann--Feynman theorem~\cite{Perdew1996} if one considers pure-state $v^\lambda$-representable densities. But the Hellmann--Feynman theorem has to be at least handled with care in the case of degeneracies\cite{Zhang2002,Balawender2004,Fernandez2004}.

\section{Differential properties of the adiabatic connection}
\label{sec:diffprop}

Note that in Eq.~\eqref{eq:F-fundamental-th} the choice of the integrand is not limited to $W(\lambda)$, but it can be any supergradient of $F(\lambda)$.
A different choice would be the right-derivative
\begin{equation}
    \rightder F(\lambda) = \lim_{\mu\searrow 0} \frac{F(\lambda+\mu) - F(\lambda)}{\mu},
\end{equation}
where the limit always exists as $F(\lambda)$ is concave.
The superdifferential of a concave function is always given as the closed interval from the right-derivative to the left-derivative, so we have for $\lambda \neq 0$
\begin{equation}\label{eq:W-from-interval}
    W(\lambda) \in [\rightder F(\lambda), \leftder F(\lambda)].
\end{equation}

\begin{figure}[ht]
\begin{tikzpicture}[scale=.65]
\draw[->,thick] (-1,0) -- (10,0) node[right]{$\lambda$};
\draw[-,thick,dashed] (0,1) -- (9.8,1) node[right]{$T(0)$};
\draw[-,thick,dashed] (0,1) -- (6,6.25) node[right]{$T(0) + \lambda W(0)$};
\draw[->,thick] (0,-0.35) -- (0,6.25);
\draw (0,1) to[out=0, in=190] (3,1.2) to[out=10, in=185] (10,3.6) node[right] {$ T(\lambda) $};
\draw (10,4.6) node[above] {$ F(\lambda) = T(\lambda) + \lambda W(\lambda)$} parabola (0,1);
\end{tikzpicture}
\caption{Illustration of the different functionals with $F(\lambda) = T(\lambda) + \lambda W(\lambda)$ concave and strictly increasing.}
\label{fig:functionals-relation}
\end{figure}
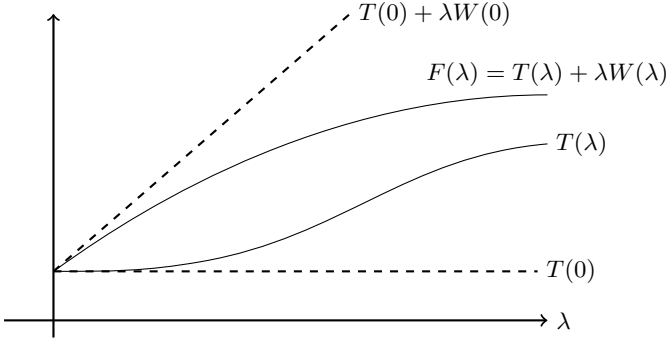

Next, we will show how the right-derivative, that will be substantial for our later definition of the exchange energy, relates to the interaction energy $W(\lambda)$.
From $F(\lambda) = T(\lambda) + \lambda W(\lambda)$ and $F(0)=T(0)$ we also have that
\begin{equation}\label{eq:F-T-frac}
    \frac{F(\lambda)-F(0)}{\lambda} - \frac{T(\lambda)-T(0)}{\lambda} = W(\lambda).
\end{equation}
The limit $\lambda \searrow 0$ is well-defined for the first term in Eq.~\eqref{eq:F-T-frac} since it equals the right-derivative of $F(\lambda)$. Since $W(\lambda)$ is monotone decreasing and finite as the subgradient of the concave and finite $F(\lambda)$, also here the limit exists and we \emph{define} $W(0) = \lim_{\lambda \searrow 0} W(\lambda)$. An alternative way to see that $W(\lambda)$ is finite is through the bound of Eq.~\eqref{eq:W-T-bound} that can be easily made into an estimate in terms of $F(\lambda)$. So far, we have defined $W(\lambda)$ only for $\lambda \neq 0$, see Eq.~\eqref{eq:W-lambda-formula} and the discussion related to it. The limit $\lambda \searrow 0$ of Eq.~\eqref{eq:F-T-frac} thus gives
\begin{equation}\label{eq:rightder-F-T-and-W}
    \rightder F(0) - \rightder T(0) = W(0).
\end{equation}
The relations between the different functionals are illustrated in Fig.~\ref{fig:functionals-relation}.

It is worth noting that for a $\rho$ that is $v$-representable and has a certain regularity, the treatment in Sec.~4 of \citet{Lammert2007} shows that $F[\rho_\gamma]$ is differentiable with respect to $\gamma$ and thus also $F(\lambda)$ for that $\rho$ is differentiable with respect to $\lambda$ for $\lambda > 0$. This would be beneficial, since then the supergradient would always be unique for $\lambda > 0$, but we will continue here without any such assumption.

In the case when $F(\lambda)$ is differentiable on at least a small interval $(0,\lambda_0)$, the supergradient is unique and it thus holds that $\rightder F(\lambda) = W(\lambda)$ for all $\lambda \in (0,\lambda_0)$ since $W(\lambda)$ was found to be a supergradient. Consequently, in this case it holds that $\lim_{\lambda\searrow 0}\rightder F(\lambda) = \rightder F(0) = W(0)$ since the right-derivative is right-continuous (§1.8 in \citet{tiel1984convex}). If $F(\lambda)$ is not differentiable, then as a concave function it only has a countable number of points $\{\lambda_i\}_i$, $0 < \lambda_{i+1} < \lambda_i < \eps$, where it is non-differentiable (again §1.8 in \citet{tiel1984convex}). If this is a finite number of points then we will just be back to the previous case, since the smallest $\lambda_i$ will then fill in the role for $\lambda_0$ from before. So we are left with an infinite sequence of points $\{\lambda_i\}_i$ that has $\lim_{i\to\infty}\lambda_i=0$ and where it holds (§1.6--7 in \citet{tiel1984convex} but for a concave function, and using that $F(\lambda)$ is non-differentiable at $\lambda_i$ and $\lambda_{i+1}$ and thus right-derivative and left-derivative cannot agree)
\begin{equation}\label{eq:der-F-inequality}
    \rightder F(\lambda_i) < \leftder F(\lambda_i) \leq \rightder F(\lambda_{i+1}) < \leftder F(\lambda_{i+1}).    
\end{equation}
The $\rightder F(\lambda_i)$, as a convergent sequence with $\rightder F(\lambda_i) \to \rightder F(0) < \infty$, must have $\rightder F(\lambda_{i+1}) - \rightder F(\lambda_i) \to 0$. But from Eq.~\eqref{eq:der-F-inequality} it then follows that also $\leftder F(\lambda_i) - \rightder F(\lambda_i) \to 0$, and so no matter how $W(\lambda_i)$ is chosen from the superdifferential of $F(\lambda)$ at $\lambda=\lambda_i$ (cf.\ Eq.~\eqref{eq:W-from-interval}), in the limit only the possibility $W(0) = \lim_{i\to\infty}W(\lambda_i) = \lim_{i\to\infty}\rightder F(\lambda_i)=\rightder F(0)$ remains. This is the main result in this section and helps to unambiguously define $W(0)$ in the given setting. From Eq.~\eqref{eq:rightder-F-T-and-W} it then follows that
\begin{equation}\label{eq:trder}
    \rightder T(0)=0
\end{equation}
and the above derived expression
\begin{equation}\label{eq:frder}
    W(0)=\lim_{\lambda\searrow 0}W(\lambda)=\rightder F(0)
\end{equation}
will yield the exchange energy, as demonstrated in the next section.

\section{Additional remarks on \texorpdfstring{$T(\lambda)$}{T(lambda)} and \texorpdfstring{$W(\lambda)$}{W(lambda)}}
\label{sec:remarks}

In this section---which can be safely skipped at first reading as the rest of the paper is independent from it---we first want to mention an alternative route for the definition of the functions $T(\lambda)$ and $W(\lambda)$ that was suggested by the Anonymous Referee that avoids any assumption of $v$-representability. 
Let $\mathfrak{G}(\lambda)$ be the set of all minimizers $\Gamma(\lambda)\mapsto\rho$ for a fixed $\rho$ in Eq.~\eqref{eq:F-def}. This set is non-empty by the result quoted before Eq.~\eqref{eq:F-Gamma}. 
Then, for example, we can make the following choice in order to define $W(\lambda)$,
\begin{equation}
    W(\lambda) = \inf_{\Gamma\in\mathfrak{G}(\lambda)} \trace{\hat W \Gamma}.
\end{equation}
This then fixes $T(\lambda)$ for all $\lambda\neq 0$ by the relation of Eq.~\eqref{eq:FTW} as 
\begin{equation}
    T(\lambda) = \sup_{\Gamma\in\mathfrak{G}(\lambda)} \trace{\hat T \Gamma}.
\end{equation}
If $\lambda=0$ then $T(0)=F(0)$ anyway.
Note that any other choice of the form
\begin{equation}\label{eq:W-infsup-interval}
    W(\lambda) \in \left[ \inf_{\Gamma\in\mathfrak{G}(\lambda)} \trace{\hat W \Gamma}, \sup_{\Gamma\in\mathfrak{G}(\lambda)} \trace{\hat W \Gamma} \right]
\end{equation}
in the interval between minimum and maximum is possible and would again fix $T(\lambda)$.

A second remark concerns Remark~3.1 in \citet{Lewin2023} which states that not only $W(\lambda)$, but also the right-derivative and the left-derivative of $F(\lambda)$, and thus by convex combination all values in-between, come from $\trace\hat W\Gamma(\lambda)$ with different minimizers $\Gamma(\lambda)$ from Eq.~\eqref{eq:F-def}. This is equivalent to saying that the interval in Eq.~\eqref{eq:W-from-interval} agrees with the one from Eq.~\eqref{eq:W-infsup-interval}. Since for $v^\lambda$-representable densities $\trace\hat W\Gamma(\lambda)$ has a fixed value for all minimizers, as was noted in Sec.~\ref{sec:coupling-strength-functionals}, this would imply that the right-derivative and left-derivative agree and thus $F(\lambda)$ would be differentiable at $\lambda$. 
Nevertheless, we have not been able to provide a proof for this statement at this point so we do not treat $F(\lambda)$ as differentiable here.

\section{Zero-coupling limit and exchange energy}
\label{sec:zero-coupling}

We now continue with our main discussion and first establish a relation between the interaction term and the exchange energy. 
In the Kohn--Sham (KS) scheme an interacting system ($\lambda>0$) and a non-interacting system ($\lambda=0$) are related by demanding that their ground-state densities are equal. This defines the Hartree-exchange-correlation energy
\begin{equation}
    \lambda\EHxc(\lambda) = F(\lambda) - F(0).
\end{equation}
It is customary to separate out the Hartree mean-field term, defined as
\begin{equation}\label{eq:Hartree}
\EHartree = \frac{1}{2}\iint \frac{\rho(\rr)\rho(\rr')}{|\rr-\rr'|}\dd\rr\dd\rr',
\end{equation}
so that one remains with just the exchange-correlation energy
\begin{equation}\label{eq:Exc-def}
    \lambda\Exc(\lambda) = F(\lambda) - F(0) - \lambda\EHartree.
\end{equation}
Using the integral formula from Eq.~\eqref{eq:F-fundamental-th}, we readily obtain
\begin{equation}
    \lambda \Exc(\lambda) = \int_{0}^\lambda W(\mu) \, \dd \mu  - \lambda\EHartree.
\end{equation}

By substituting $\lambda = 1/\gamma > 0$ and by using the scaling relation of Eq.~\eqref{eq:F-scaling} we have
\begin{equation}
    \begin{aligned}
         \Exc( 1/ \gamma ) &= \frac{F( 1/ \gamma ) - F(0)}{1/\gamma} - \EHartree \\
         & =  \frac 1 \gamma \left( \gamma^2\left(  F( 1/ \gamma ) - F(0) \right)  - \gamma \EHartree \right)\\
         & =  \frac 1 \gamma \left( F^{\lambda=1}[\rho_\gamma] - T^{\lambda=0}[\rho_\gamma] - \EHartree[\rho_\gamma]   \right) \\
         & =  \frac 1 \gamma \Exc[\rho_\gamma]
    \end{aligned}
\end{equation}
and thus in the limit using Eq.~\eqref{eq:frder},
\begin{equation}\label{eq:Ex-from-limit}
\begin{aligned}
    \Ex &= \lim_{\gamma \to \infty} \frac 1 \gamma \Exc[\rho_\gamma] 
    =  \lim_{\lambda \searrow 0} \Exc( \lambda ) \\
    &= \lim_{\lambda \searrow 0} \frac{F( \lambda ) - F(0)}{\lambda} - \EHartree \\
    &= \rightder F(0) - \EHartree = W(0) - \EHartree.
\end{aligned}
\end{equation}
This coincides exactly with definition in \citet{LevyPerdewPhysRevA1985} and beautifully relates the definition of the exchange energy as the high-density limit through coordinate scaling with its alternative definition as a zero-coupling limit. It further yields a third definition  of the exchange energy as the right-derivative of the Lieb functional at zero coupling strength minus the Hartree energy. This also shows that if we split the exchange-correlation energy in the usual way,
\begin{equation}
    \Exc(\lambda) = \Ex + \Ec(\lambda),
\end{equation}
then $\Ec(\lambda)$ carries all $\lambda$ dependence of $\Exc(\lambda)$, or, said differently, the linear (in $\lambda$) contribution to $\lambda \Exc(\lambda)$ is exactly equal to $\Ex$ and at zero coupling it holds $\Ec(0) = 0$. This fact can be taken as a motivation to define $\Ex$ by Eq.~\eqref{eq:Ex-from-limit} instead.

Note that in the presented setting, the exchange-correlation potential for different coupling strengths $\lambda$ can be defined through 
(by assumption of $v^\lambda$-representability)
\begin{equation} \label{eq:pot-lambda-split}
    v^0(\rr) = v^\lambda(\rr) + \lambda \vHxc^\lambda(\rr) = 
    v^\lambda(\rr) + \lambda \vH(\rr) + \lambda \vxc^\lambda(\rr),
\end{equation}
where 
\begin{equation}
\vH(\rr) = \int \frac{\rho(\rr')}{|\rr-\rr'|} \dd \rr'
\end{equation}
is the usual Hartree potential. In analogy to Eq.~\eqref{eq:Ex-from-limit} it is then tempting to define the local-exchange potential as the zero-coupling limit $\vx(\rr) = \lim_{\lambda\searrow 0} \vxc^\lambda(\rr)$ (since the usual definition as the functional derivative of $\Ex$ needs to be avoided) and introduce the split of the exchange-correlation potential $\vxc^\lambda(\rr) = \vx(\rr) + \vc^\lambda(\rr)$. However, we cannot be sure about the existence of the zero-coupling limit for the xc potential. In the next section, we are only able to establish the limits for the energy expressions in the context of the virial relation.

\section{Exchange-only virial relation}
\label{sec:xVR}

In this section, we present our main result about the exchange-only virial relation (Eq.~\eqref{eq:Ex-VR-main-result} from Sec.~\ref{sec:intro}, or Eq.~\eqref{eq:Ex-VR-new} below). 
By using the usual virial relation from Eq.~\eqref{eq:full-VR} we arrive easily at the form given by \citet{LevyPerdewPhysRevA1985} for $\lambda > 0$ and $\lambda = 0$,
\begin{align}
    &F(\lambda) + T(\lambda) = \int \rho(\rr) \,\rr \cdot \nabla v^\lambda(\rr) \,\dd \rr,\\
    &F(0) + T(0) = \int \rho(\rr) \,\rr \cdot \nabla v^0(\rr) \,\dd \rr.
\end{align}
We subtract those and divide by $\lambda \neq 0$ to get
\begin{equation}\label{eq:VR-diff}
\begin{aligned}
    \frac{1}{\lambda}(F(\lambda)&-F(0)) + \frac{1}{\lambda}(T(\lambda)-T(0)) \\
    &= - \int \rho(\rr) \,\rr \cdot \nabla \vHxc^\lambda(\rr) \,\dd \rr.
\end{aligned}
\end{equation}
Here, we introduced the ($\lambda$- and $\rho$-dependent) Hartree-exchange-correlation potential $\lambda\vHxc^\lambda(\rr) = v^0(\rr) - v^\lambda(\rr)$, which of course relies on $v^\lambda$-representability again. In the limit $\lambda\searrow 0$, we then obtain that
\begin{equation}\label{eq:zero-limit}
    \rightder F(0) + \rightder T(0) = - \lim_{\lambda\searrow 0}\int \rho(\rr) \,\rr \cdot \nabla \vHxc^\lambda(\rr) \,\dd \rr.
\end{equation}
Using Eq.~\eqref{eq:Ex-from-limit}, Eq.~\eqref{eq:trder}, and additionally the virial relation for the Hartree term that follows from direct computation,
\begin{equation}
    \EHartree = -\int \rho(\rr) \,\rr \cdot \nabla \vH(\rr) \,\dd \rr,
\end{equation}
that we have with $\vHxc^\lambda(\rr) = \vH(\rr) + \vxc^\lambda(\rr)$ the exchange-only virial relation
\begin{equation}\label{eq:Ex-VR-new}
    \Ex = -\lim_{\lambda\searrow 0}\int \rho(\rr) \,\rr \cdot \nabla \vxc^\lambda(\rr) \,\dd \rr.
\end{equation}
It is interesting to contrast this with the result from the force-based treatment in \citet{other-special-edition-paper}, where the virial relation includes an additional transversal term next to the gradient of the force-based local exchange potential $v_{\rm fx}$,
\begin{equation}
    \Ex = \int \rho(\rr) \, \rr \cdot (-\nabla v_{\rm fx}(\rr) + \nabla \times \boldsymbol{\alpha}_{\rm fx}(\rr)) \,\dd\rr.
\end{equation}
The necessary appearance of this additional term (within the integrand), which will be non-zero in all non-trivial situations, already makes it clear that we cannot expect $v_{\rm fx}(\rr) = \lim_{\lambda\searrow 0} \vxc^\lambda(\rr)$. 
We can, however, note that $\int \rho(\rr) \, \rr \cdot  \nabla \times \boldsymbol{\alpha}_{\rm fx}(\rr) \dd\rr =0$ for spherically-symmetric systems~\cite{other-special-edition-paper}, which implies for such systems an equality between $\lim_{\lambda\searrow 0}\int \rho(\rr) \,\rr \cdot \nabla \vxc^\lambda(\rr) \dd \rr$ and $ \int \rho(\rr) \, \rr \cdot \nabla v_{\rm fx}(\rr) \dd\rr$. 
Further results connecting to the force-based approach beyond this obvious observation is relegated to future work.

\section{Summary and Discussion}

The adiabatic connection relates the different energy contributions of a quantum many-body system in its ground state with variable coupling strength $\lambda$ and fixed one-particle density $\rho$.
Central to our work is the assumption of $v^\lambda$-representability of $\rho$, which means that $\rho$ is $v$-representable at each coupling strength $\lambda \geq 0$.  
This allowed us assign unique values to the kinetic-energy function $T(\lambda)$ and the interaction-energy function $ W(\lambda)$, see Sec.~\ref{sec:coupling-strength-functionals}. Various convex-analytic properties of the adiabatic connection were studied in Sec.~\ref{sec:prop} and Sec.~\ref{sec:diffprop}. We found that $W(\lambda)$ is exactly a supergradient of $F(\lambda)$ at $\lambda\neq 0$ and if defined as $W(0) = \lim_{\lambda\searrow 0}W(\lambda)$ at $\lambda=0$ then $W(0)$ is given by the right-derivative $\rightder F(0)$ of $F(\lambda)$ at $\lambda=0$. 
We then confirmed the usual relation $\Ex = W(0) - \EHartree$, which now alternatively allows a definition of the exchange energy through the adiabatic connection as $ \rightder F(0) - \EHartree = \Ex$. 
The last equation connects the zero-coupling limit to the high-density limit from coordinate scaling, since $\Ex$ is here defined through the latter.   

In this context, it is natural to also ask for the virial relation~\eqref{eq:Ex-VR} relating the exchange energy with an exchange potential and we examined this possibility in Sec.~\ref{sec:xVR}. 
While the zero-coupling limits in Eq.~\eqref{eq:zero-limit} can indeed be guaranteed to exist, this is unlikely to be the case for the associated exchange-correlation potential $\vxc^\lambda$ without further assumptions. 
Although one would usually like to set $\vx = \lim_{\lambda\searrow 0}\vxc^\lambda $, 
only the limit of the corresponding virial integral in Eq.~\eqref{eq:Ex-VR-new} is proven to exist to the exchange energy. This means that while we have a well-defined exchange energy $\Ex[\rho]$ for a $v^\lambda$-representable density $\rho$ from Eq.~\eqref{eq:Ex-from-limit}, there might not be a general well-defined local-exchange potential from a similar limit or a functional derivative. This realization is in line with the results from \citet{other-special-edition-paper} in the same special issue of this journal. 
In that work, following a definition of the exchange contribution in terms of force densities, one needs an additional vector potential to fulfill the exchange-only virial relation~\eqref{eq:Ex-VR}. That the exchange-only virial relation does not hold in general can also be seen from numerical evidence~\cite{other-special-edition-paper}. Of course, techniques like the optimized-effective potential (OEP) method~\cite{SHARP_PR90_317,TALMAN_PRA14_36,YangWu2002} can still be used to get an approximate local-exchange potential.

\begin{acknowledgements}
AL, MAC and MP have received funding from the ERC-2021-STG under grant agreement No.~101041487 REGAL.
AR, NT-D and MR was supported by the European
Research Council (ERC-2015-AdG694097) and by the Cluster of
Excellence ``CUI: Advanced Imaging of Matter'' of the Deutsche
Forschungsgemeinschaft (DFG) -- EXC 2056 -- project ID 390715994,
and the Grupos Consolidados (IT1249-19).
TH was supported by the Research Council of Norway through “Magnetic Chemistry” Grant
No.~287950. AL and MAC have received funding from the Research Council of Norway through CCerror Grant No.~287906.
AL, MAC and TH were funded by the Research Council of Norway through CoE Hylleraas Centre for Quantum Molecular Sciences Grant No.~262695.

\end{acknowledgements}

%

\end{document}